\documentclass[rsl]{ursi}
\usepackage{amsfonts}
\usepackage{tikz}
\usepackage{pgfplots}
\usetikzlibrary{calc}
\pgfplotsset{compat=1.18}
\usetikzlibrary{intersections}
\usepgfplotslibrary{fillbetween}

\makeatletter

\def\blfootnote{\gdef\@thefnmark{}\@footnotetext}

\makeatother

\title{A Transformer for Joint Multi-Receiver Pilotless Wi-Fi Decoding}

\author{Xavier Tardy*\affref{ref1}\affref{ref2}, Grégoire Lefebvre\affref{ref2},
  Apostolos Kountouris\affref{ref2}, Haïfa Farès\affref{ref1}, and Amor Nafkha\affref{ref1}}

\affiliation{%
  \aff{ref1}{IETR - UMR CNRS 6164, CentraleSupélec, Cesson-Sévigné, France}
  \aff{ref2}{Orange Research, Grenoble, France}
}

\begin{document}

\maketitle

\begin{abstract}
In this paper we discuss the development of a fully pilotless multi-access-point Wi-Fi receiver. This is based on a self-attention Transformer that operates on per-(access point, subcarrier) tokens and outputs bit-wise logits for a standard low-density parity-check decoder. 
In realistic ray-traced indoor channels, pilotless decoding becomes feasible as soon as some spatial diversity is available. While a single-access point configuration does not attain practical error rates, when compared to an ideal maximum-ratio-combining baseline with perfect channel knowledge, cooperative configurations with three access points achieve coded bit error rates of approximately $10^{-5}$. This is achieved while benefiting from higher spectral efficiency by eliminating pilots.
\end{abstract}

\section{Introduction}

\blfootnote{This work was supported in part by Bpifrance under the France 2030 i‑Démo program (Wi‑FIP project, 2023–2026, Grant I-DEMO-52255).}

Future Wi-Fi generations (e.g., IEEE 802.11be/bn)~\cite{ref_wifi} aim to achieve high spectral efficiency in dense deployments through multi-link coordination. Although these standards do not explicitly envision fully pilotless operation, further reduction of pilot overhead remains an attractive option, especially for the uplink in such dense, coordinated environments. In these scenarios, multiple access points (APs) observe the same transmission, and reusing any pilot resource as data can directly enhance the net rate on that specific link. While removing pilots helps reduce overhead, it also introduces challenges for decoding in frequency-selective indoor channels. Without explicit channel state information (CSI), the receiver must contend with amplitude and phase distortions, as well as deep fades.

To address these challenges, recent approaches have explored end-to-end learned pilotless schemes~\cite{ref_e2e_ofdm,ref_trim_fat} and have shown promising results. However, these schemes typically rely on a jointly trainable transmitter and receiver which is a configuration that is difficult to reconcile with standardized Wi-Fi waveforms. In practice, coordinated multi-AP reception offers spatial diversity that can mitigate deep fades and stabilize demapping. When multiple geographically separated APs observe the same uplink symbol, they can be jointly processed at a central unit to improve reliability. Traditional pilot-based pipelines are not ideally suited for environments lacking dedicated pilots because they first estimate per-AP channels and then combine soft metrics such as maximum ratio combining (MRC). These conventional methods do not fully exploit the frequency-selective reliability differences across APs, leaving room for the consideration of more advanced processing techniques. 

In this study these challenges are addressed with a full self-attention Transformer~\cite{ref_transformer} that consumes raw complex frequency-domain observations from multiple APs and outputs bit-wise logits. This avoids the need for explicit CSI and dedicated pilots while keeping a standard Wi‑Fi-compatible QAM waveform. It also enables practical pilotless multi-AP decoding with a purely receiver-side neural design.

\section{Related Work}

Conventional Wi‑Fi and orthogonal frequency-division multiplexing (OFDM) receivers rely on pilot-based channel estimation, followed by per-subcarrier equalization and soft demapping. With dense pilots, least-squares (LS) or linear minimum mean-square error (LMMSE) estimators~\cite{ref_ls_lmmse_ofdm} provide accurate CSI but at the cost of spectral overhead and strong degradation when pilots are sparse or removed, especially in indoor frequency-selective channels. In coordinated multi-AP reception, such pipelines are usually run independently per AP and combined a posteriori, 
which ignores frequency-selective reliability differences and underutilizes cooperative gains~\cite{yilmaz_alouini_mgf_2012}.

Consequently, data-driven receivers have been proposed to mitigate dependence on explicit CSI and decrease the need for pilots. End-to-end learned OFDM links demonstrate that neural receivers can operate with superimposed or no dedicated pilots while preserving the bit error rate (BER) compared to classical baselines~\cite{ref_e2e_ofdm,ref_trim_fat}, but they jointly train the transmitter and receiver (e.g., learned constellations or superimposed pilots), which is challenging to deploy in standardized Wi‑Fi where the transmitter is not under control. Other works replace parts of the model-based chain with CNN- or Transformer-based receivers operating on the time–frequency grid 
and output soft bits for a conventional channel decoder~\cite{honkala_deeprx_2021,xie_comm-transformer_2024}; nevertheless, the authors focus on single-receiver topologies and assume some pilot structure.

In contrast, the present study targets a pilotless multi-AP Wi‑Fi uplink with a conventional standardized transmitter. Instead of estimating and combining per-AP channels, a self-attention Transformer processes raw complex frequency-domain samples from all APs and outputs bit-wise logits for a standard low-density parity-check (LDPC) decoder. This process facilitates learning content-aware fusion across APs and wideband processing across subcarriers without explicit CSI.

\section{System Model and Problem Statement}

The system model considers a single-antenna uplink user equipment (UE) transmitting a coded OFDM symbol to $N_R$ single-antenna APs. A binary LDPC code of rate $R=2/3$ encodes the information bits, which are mapped onto an $M$-QAM constellation and placed on $N_c$ active subcarriers of a single OFDM symbol, yielding the transmit vector $\mathbf{x}\in\mathbb{C}^{N_c}$. We focus on per-symbol processing, assuming ideal time/frequency synchronization (or residual offsets absorbed into the effective channel) and an ideal low-latency fronthaul to a centralized processor.

For each AP $r\in\{1,\dots,N_R\}$, the received frequency-domain symbol is defined as follows:
\begin{equation}
  \mathbf{y}^{(r)} = \mathbf{h}^{(r)} \circ \mathbf{x} + \mathbf{n}^{(r)}, \quad
  \mathbf{n}^{(r)} \sim \mathcal{CN}(\mathbf{0},\sigma_r^2\mathbf{I}),
  \label{eq:rx}
\end{equation}
where $\mathbf{h}^{(r)}\in\mathbb{C}^{N_c}$ is the effective per-subcarrier channel and $\circ$ denotes element-wise multiplication. Stacking the received vectors from all APs into the rows of a matrix yields
\begin{equation}
  \mathbf{Y}
  =
  \begin{bmatrix}
    \mathbf{y}^{(1)} & \cdots & \mathbf{y}^{(N_R)}
  \end{bmatrix}^{\!\mathsf{T}}
  \in \mathbb{C}^{N_R\times N_c},
\end{equation}



Let $\mathbf{c}\in\{0,1\}^{mN_c}$ denote the coded bits carried by the $N_c$ QAM symbols of the considered OFDM symbol, where $m$ is the number of bits per symbol. The goal is to learn a mapping $g_{\theta}:\big(\mathbf{Y},\{\sigma_r^2\}_{r=1}^{N_R}\big)\mapsto\mathbf{LLR}\in\mathbb{R}^{mN_c}$ with a Transformer $g_{\theta}$ that directly outputs bit-wise logits, which are log-likelihood-ratio (LLR) compatible for $\mathbf{c}$, from the raw multi-AP observations without explicit CSI. 

The Transformer $g_{\theta}$ is trained by minimizing the bit-wise binary cross-entropy (BCE), defined as:
\begin{equation}
  \mathcal{L}(\theta)
  = \frac{1}{mN_c}\sum_{i=1}^{mN_c}
  \log\!\big(1+\mathrm{e}^{-s_i\,\mathbf{LLR}_{i}}\big),
  \quad
  s_i = 2c_{i}-1.
  \label{eq:bce}
\end{equation}
This strategy is equivalent (up to a constant) to maximizing the bit-metric decoding rate. At test time, $\mathbf{LLR}$ is passed to a conventional soft-input LDPC decoder to assess the coded BER, while hard decisions on $\mathbf{LLR}$ provide the uncoded BER.

\section{Full Self-Attention Receiver}

\subsection{Per-(AP, Subcarrier) Tokens}

A single OFDM symbol is processed with $N_c$ subcarriers and $N_R$ receivers. For each AP $r$ and subcarrier $f$, the received sample is $Y^{(r)}_f = \mathbf{Y}_{r,f}$. We build one token per (AP, subcarrier) pair with a compact real-valued feature vector, defined as:
\begin{equation}
  \mathbf{u}^{(r)}_f
  =
  \big[
    \operatorname{Re}\{Y^{(r)}_f\},
    \ \operatorname{Im}\{Y^{(r)}_f\},
    \ \sigma_r^2
  \big]^{\mathsf{T}}
  \in \mathbb{R}^{3},
  \label{eq:u_rf}
\end{equation}
where $\sigma_r^2$ is the noise variance estimate at AP~$r$. The total number of tokens is thus $N_{\text{token}} = N_R \times N_c$.
Each token represents the observation of one subcarrier at one AP.

\subsection{Embedding and Self-Attention Encoder}

Each token is linearly embedded into the model dimension $d_{\text{model}}$, as follows:
\begin{equation}
  \mathbf{z}^{(r)}_{0,f}
  =
  \mathbf{W}_e\,\mathbf{u}^{(r)}_f + \mathbf{b}_e
  \in \mathbb{R}^{d_{\text{model}}},
\end{equation}
with shared weight $\mathbf{W}_e \in \mathbb{R}^{d_{\text{model}}\times 3}$ and bias $\mathbf{b}_e \in \mathbb{R}^{d_{\text{model}}}$ parameters. We add a fixed sinusoidal positional embedding that depends only on the subcarrier index $f$, defined as:
\begin{equation}
  \tilde{\mathbf{z}}^{(r)}_{0,f}
  =
  \mathbf{z}^{(r)}_{0,f}
  +
  \mathbf{e}^{(\mathrm{sc})}_f,
  \qquad
  \mathbf{e}^{(\mathrm{sc})}_f \in \mathbb{R}^{d_{\text{model}}},
\end{equation}
where the vectors $\mathbf{e}^{(\mathrm{sc})}_f$ follow the standard sinusoidal positional encoding scheme as in~\cite{ref_transformer}. This allows tokens sharing the same subcarrier but different APs to be distinguished while exposing the frequency structure to the network. Stacking all $(r,f)$ tokens yields
$  \mathbf{Z}_0 \in \mathbb{R}^{N_{\text{token}}\times d_{\text{model}}}$.
We then apply a standard stack of $4$ Transformer encoder layers~\cite{ref_transformer} to $\mathbf{Z}_0$. Each layer consists of multi-head self-attention over the full set of $N_{\text{token}}$ tokens, followed by a position-wise feed-forward network, with residual connections and layer normalization. This produces
$  \mathbf{Z}_L \in \mathbb{R}^{N_{\text{token}}\times d_{\text{model}}}$,
a context-aware representation for each (AP, subcarrier) token that captures both wideband correlations across subcarriers and spatial correlations across APs.

\subsection{Subcarrier Fusion and LLR Prediction}

$\mathbf{Z}_L$ is reshaped back into $(r,f)$ indices, yielding $\mathbf{z}^{(r)}_{L,f}$ for all receivers and subcarriers. For each subcarrier $f$, the $N_R$ AP embeddings are fused through averaging, defined as:
\begin{equation}
  \bar{\mathbf{z}}_f
  =
  \frac{1}{N_R}
  \sum_{r=1}^{N_R}
  \mathbf{z}^{(r)}_{L,f}
  \in \mathbb{R}^{d_{\text{model}}},
\end{equation}
which lets the self-attention layers learn how information is shared across APs, while the final fusion remains lightweight and permutation-invariant in the AP index.

A multi-layer perceptron (MLP) maps the fused representation of subcarrier $f$
to $m$ logits (LLR-compatible) associated with the $m$ bits of the $M$-QAM symbol.
\begin{equation}
  \mathbf{LLR}_f
  =
  \mathrm{MLP}(\bar{\mathbf{z}}_f)
  \in \mathbb{R}^{m},\quad f=1,\ldots,N_c.
\end{equation}
Stacking all subcarriers gives the full LLR vector
  $\mathbf{LLR}
  =
  \big[
    \mathbf{LLR}_1^{\mathsf{T}},
    \dots,
    \mathbf{LLR}_{N_c}^{\mathsf{T}}
  \big]^{\mathsf{T}}
  \in \mathbb{R}^{mN_c}$,
which is used in the BCE loss~\eqref{eq:bce} during training and as soft input to the external LDPC decoder during testing.

\section{Experiments}

The proposed receiver is evaluated in a ray-traced indoor environment with $N_c=72$ subcarriers, 16-QAM modulation, and a cooperation level $N_R\in\{1,\dots,5\}$. During training, the Transformer is optimized on the \emph{uncoded} bit labels using the BCE loss in \eqref{eq:bce}, without any LDPC decoding in the loop, so that it does not adapt to a specific channel code. At test time, the coded BER is obtained by feeding the predicted logits into a standard rate-$2/3$ LDPC decoder and is compared to a perfect-CSI benchmark described below.

\subsection{Ray-Traced 3D Indoor Environment}

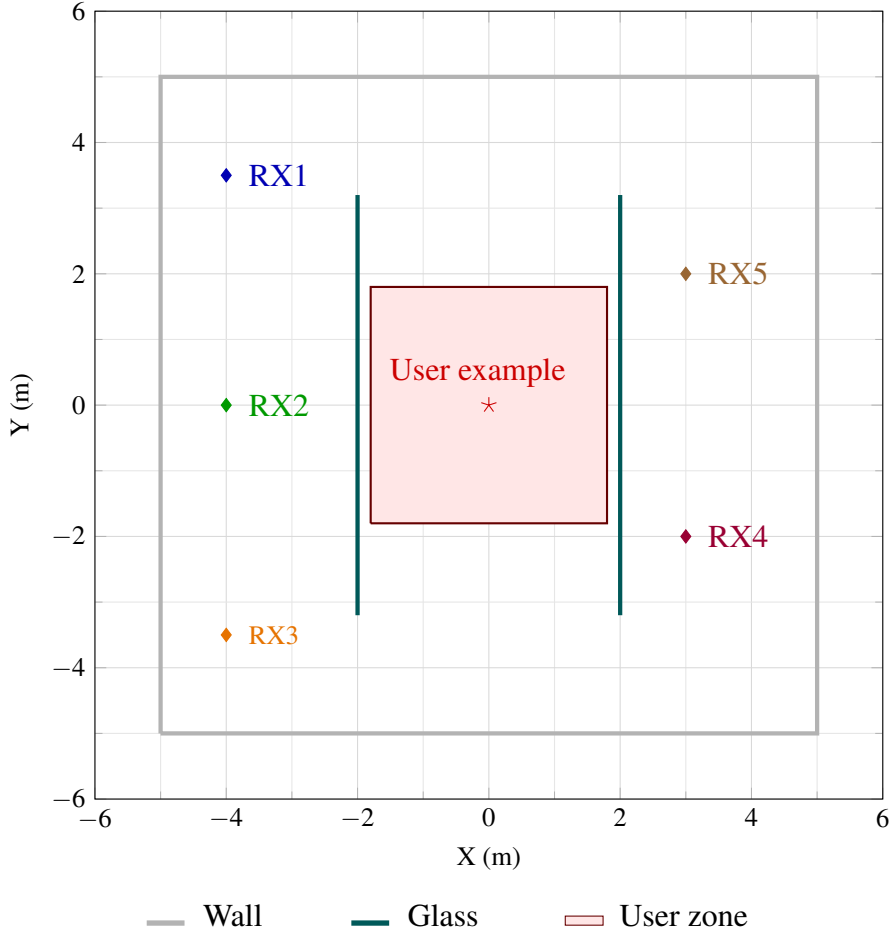
\begin{figure}[t]
    \centering
    \begin{tikzpicture}
    \centering
      \begin{axis}[
        width=\linewidth,
        height=0.75\linewidth,
        axis equal image,
        xmin=-6, xmax=6,
        ymin=-6, ymax=6,
        xlabel={X (m)}, ylabel={Y (m)},
        label style={font=\normalsize},          
        tick label style={font=\normalsize},     
        tick style={gray!60},
        grid=both,
        minor tick num=1,
        grid style={gray!20},
        major grid style={gray!30},
        enlargelimits=false,
      ]

        \addplot [gray!60, line width=1.6pt]
          coordinates {(-5,-5) (5,-5) (5,5) (-5,5) (-5,-5)};

        \addplot [teal!70!black, line width=1.6pt]
          coordinates {(-2,-3.2) (-2,3.2)};
        \addplot [teal!70!black, line width=1.6pt]
          coordinates {( 2,-3.2) ( 2,3.2)};

        \addplot [draw=red!40!black, fill=red!10, line width=0.8pt]
          coordinates {(-1.8,-1.8) (1.8,-1.8) (1.8,1.8) (-1.8,1.8) (-1.8,-1.8)};

        \addplot[
          only marks,
          mark=star,
          mark size=3pt,
          color=red!80!black
        ] coordinates {(0.0, 0.0)};
        \node[anchor=west, text=red!80!black, font=\large]
          at (axis cs:-1.65, 0.5) {User example};

        \addplot[
          only marks,
          mark=diamond*,
          mark size=2.5pt,
          color=blue!70!black
        ] coordinates {(-4.0, 3.5)};
        \node[anchor=west, font=\large, text=blue!70!black]
          at (axis cs:-3.8, 3.5) {RX1};

        \addplot[
          only marks,
          mark=diamond*,
          mark size=2.5pt,
          color=green!60!black
        ] coordinates {(-4.0, 0.0)};
        \node[anchor=west, font=\large, text=green!60!black]
          at (axis cs:-3.8, 0.0) {RX2};

        \addplot[
          only marks,
          mark=diamond*,
          mark size=2.5pt,
          color=orange!90!black
        ] coordinates {(-4.0, -3.5)};
        \node[anchor=west, font=\normalsize, text=orange!90!black]
          at (axis cs:-3.8, -3.5) {RX3};

        \addplot[
          only marks,
          mark=diamond*,
          mark size=2.5pt,
          color=purple!80!black
        ] coordinates {(3.0, -2.0)};
        \node[anchor=west, font=\large, text=purple!80!black]
          at (axis cs:3.2, -2.0) {RX4};

        \addplot[
          only marks,
          mark=diamond*,
          mark size=2.5pt,
          color=brown!80!black
        ] coordinates {(3.0, 2.0)};
        \node[anchor=west, font=\large, text=brown!80!black]
          at (axis cs:3.2, 2.0) {RX5};

      \end{axis}
    \end{tikzpicture}

  \vspace{2mm}

  \makebox[\linewidth][c]{%
    \large
    \tikz{\draw[gray!60,line width=2.0pt] (0,0) -- (0.5,0);}
    \ Wall
    \hspace{2.2em}
    \tikz{\draw[teal!70!black,line width=2.0pt] (0,0) -- (0.5,0);}
    \ Glass
    \hspace{2.2em}
    \tikz{\path[
      draw=red!40!black,
      fill=red!10
    ] (0,-0.06) rectangle (0.5,0.06);}
    \ User zone
  }

  \vspace{1mm}

  \caption{Top-down view of the 3D indoor environment used for ray-traced channel generation.}
  \label{fig:env}
\end{figure}

Channels are generated with Sionna RT~\cite{ref_sionna} in a ``Box--Two--Screens'' indoor environment. This setup mimics, for example, a Wi‑Fi deployment in a corridor or office floor with several APs, where a user moves within a central hotspot area. A single user is located in a central zone, and up to five APs are placed at distinct positions in the room on both sides of two glass partitions, as illustrated in Fig.~\ref{fig:env}. The geometry includes walls and glass screens; Sionna RT accounts for specular reflections and diffraction. Using a ray‑traced channel model is essential here to capture realistic frequency‑selective fading patterns across the different APs, which strongly impact the potential gains of cooperative reception.

\subsection{Training and Evaluation Protocol}

For each value of $N_R\in\{1,\dots,5\}$, $100{,}000$ independent channel realizations are generated in the environment of Fig.~\ref{fig:env}. Each realization corresponds to a new random user position within the central user zone, while the $N_R$ APs remain fixed. The dataset is divided into three parts: 20\% is allocated for training, 10\% for validation, and 70\% for testing. This results in a test set comprising 20,720,000 bits, which is necessary to obtain reliable BER estimates. Training is performed with a batch size of $32$, the Adam optimizer, and a learning rate of $10^{-3}$.

As a baseline, an idealized single-link OFDM receiver is simulated with perfect CSI and soft demapping. For each subcarrier, a conventional Wi-Fi link is considered with $N_R = 1$ receiving antenna, assuming perfect knowledge of the channel coefficient, and computing symbol-wise LLRs for 16-QAM, which are passed to the same LDPC decoder.

\subsection{Model Size and Complexity}

In our experiments, we design a compact Transformer encoder with $4$ layers,
a model dimension $d_{\text{model}}=64$, and $4$ attention heads per layer.
Each layer contains a position-wise feed-forward network of dimension $128$,
and the per-subcarrier prediction head is a two-layer MLP with hidden size $128$
mapping $\bar{\mathbf{z}}_f$ to $m$ logits. Overall, the receiver-side neural
model has approximately $1.35\times 10^5$ trainable parameters. A full
self-attention layer has complexity
$\mathcal{O}(N_{\text{token}}^2 d_{\text{model}})$, which is higher than that of
classical model-based receivers with linear complexity in $N_R$ and $N_c$.

\section{Results}

\begin{figure}[htbp]
  \centering
  \begin{tikzpicture}
    \begin{axis}[
      width=0.6\columnwidth,
      height=0.40\columnwidth,
      grid=both,
      grid style={gray!20},
      major grid style={gray!40},
      xmin=-2, xmax=14,
      ymin=1e-6, ymax=1,
      xmode=linear,
      ymode=log,
      xlabel={$E_b/N_0$ (dB)},
      ylabel={Coded BER},
      label style={font=\normalsize},
      tick label style={font=\normalsize},
      legend style={
        font=\normalsize,
        cells={anchor=east},
        at={(1.4,0.9)},
        anchor=north,
        draw=none,
        fill=white,
        fill opacity=0.9,
        text opacity=1,
        legend columns=1,
        row sep=8pt, 
      }
    ]

\addplot[
  semithick,
  color=blue,
  mark=o,
  mark options={scale=0.9},
  forget plot
] coordinates {
  (-2, 3.5510e-01)
  ( 0, 3.2355e-01)
  ( 2, 2.8993e-01)
  ( 4, 2.5079e-01)
  ( 6, 1.9971e-01)
  ( 8, 1.3376e-01)
  (10, 7.9021e-02)
  (12, 4.6932e-02)
  (14, 3.0434e-02)
};
\addplot[name path=upperN1, draw=none, forget plot] coordinates {
  (-2, 3.5510e-01 + 6.6817e-04)
  ( 0, 3.2355e-01 + 7.6276e-04)
  ( 2, 2.8993e-01 + 8.9654e-04)
  ( 4, 2.5079e-01 + 1.3834e-03)
  ( 6, 1.9971e-01 + 2.3292e-03)
  ( 8, 1.3376e-01 + 3.0127e-03)
  (10, 7.9021e-02 + 2.8219e-03)
  (12, 4.6932e-02 + 2.4236e-03)
  (14, 3.0434e-02 + 1.7563e-03)
};
\addplot[name path=lowerN1, draw=none, forget plot] coordinates {
  (-2, 3.5510e-01 - 6.6817e-04)
  ( 0, 3.2355e-01 - 7.6276e-04)
  ( 2, 2.8993e-01 - 8.9654e-04)
  ( 4, 2.5079e-01 - 1.3834e-03)
  ( 6, 1.9971e-01 - 2.3292e-03)
  ( 8, 1.3376e-01 - 3.0127e-03)
  (10, 7.9021e-02 - 2.8219e-03)
  (12, 4.6932e-02 - 2.4236e-03)
  (14, 3.0434e-02 - 1.7563e-03)
};
\addplot[blue!50, opacity=0.6, forget plot] fill between[of=upperN1 and lowerN1];

\addplot[
  semithick,
  color=red,
  mark=triangle*,
  mark options={scale=0.9},
  forget plot
] coordinates {
  (-2, 2.7469e-01)
  ( 0, 2.2395e-01)
  ( 2, 1.4702e-01)
  ( 4, 5.7683e-02)
  ( 6, 1.2770e-02)
  ( 8, 1.8044e-03)
  (10, 2.3586e-04)
  (12, 5.8358e-05)
  (14, 2.6451e-05)
};
\addplot[name path=upperN2, draw=none, forget plot] coordinates {
  (-2, 2.7469e-01 + 7.7676e-04)
  ( 0, 2.2395e-01 + 9.2072e-04)
  ( 2, 1.4702e-01 + 1.5484e-03)
  ( 4, 5.7683e-02 + 1.6188e-03)
  ( 6, 1.2770e-02 + 7.7858e-04)
  ( 8, 1.8044e-03 + 2.0766e-04)
  (10, 2.3586e-04 + 5.2499e-05)
  (12, 5.8358e-05 + 2.5954e-05)
  (14, 2.6451e-05 + 1.4116e-05)
};
\addplot[name path=lowerN2, draw=none, forget plot] coordinates {
  (-2, 2.7469e-01 - 7.7676e-04)
  ( 0, 2.2395e-01 - 9.2072e-04)
  ( 2, 1.4702e-01 - 1.5484e-03)
  ( 4, 5.7683e-02 - 1.6188e-03)
  ( 6, 1.2770e-02 - 7.7858e-04)
  ( 8, 1.8044e-03 - 2.0766e-04)
  (10, 2.3586e-04 - 5.2499e-05)
  (12, 5.8358e-05 - 2.5954e-05)
  (14, 2.6451e-05 - 1.4116e-05)
};
\addplot[red!50, opacity=0.6, forget plot] fill between[of=upperN2 and lowerN2];

\addplot[
  semithick,
  color=green!70!black,
  mark=square*,
  mark options={scale=0.9},
  forget plot
] coordinates {
  (-2, 2.1992e-01)
  ( 0, 1.3535e-01)
  ( 2, 3.7346e-02)
  ( 4, 5.1150e-03)
  ( 6, 4.6798e-04)
  ( 8, 2.6079e-05)
  (10, 1.5997e-06)
  (12, 2.6042e-07)
  (14, 0.0000e+00)
};
\addplot[name path=upperN3, draw=none, forget plot] coordinates {
  (-2, 2.1992e-01 + 9.1950e-04)
  ( 0, 1.3535e-01 + 1.9521e-03)
  ( 2, 3.7346e-02 + 1.6554e-03)
  ( 4, 5.1150e-03 + 3.5203e-04)
  ( 6, 4.6798e-04 + 5.0579e-05)
  ( 8, 2.6079e-05 + 3.1089e-06)
  (10, 1.5997e-06 + 1.3106e-06)
  (12, 2.6042e-07 + 5.8231e-07)
  (14, 0.0000e+00 + 0.0000e+00)
};
\addplot[name path=lowerN3, draw=none, forget plot] coordinates {
  (-2, 2.1992e-01 - 9.1950e-04)
  ( 0, 1.3535e-01 - 1.9521e-03)
  ( 2, 3.7346e-02 - 1.6554e-03)
  ( 4, 5.1150e-03 - 3.5203e-04)
  ( 6, 4.6798e-04 - 5.0579e-05)
  ( 8, 2.6079e-05 - 3.1089e-06)
  (10, 1.5997e-06 - 1.3106e-06)
  (12, 2.6042e-07 - 5.8231e-07)
  (14, 0.0000e+00 - 0.0000e+00)
};
\addplot[green!70!black!50, opacity=0.6, forget plot] fill between[of=upperN3 and lowerN3];

\addplot[
  semithick,
  color=orange!90!black,
  mark=diamond*,
  mark options={scale=0.9},
  forget plot
] coordinates {
  (-2, 1.6772e-01)
  ( 0, 5.8392e-02)
  ( 2, 5.9270e-03)
  ( 4, 1.9771e-04)
  ( 6, 8.8541e-06)
  ( 8, 7.6885e-07)
  (10, 4.8363e-07)
  (12, 0.0000e+00)
  (14, 0.0000e+00)
};
\addplot[name path=upperN4, draw=none, forget plot] coordinates {
  (-2, 1.6772e-01 + 1.9573e-03)
  ( 0, 5.8392e-02 + 2.1831e-03)
  ( 2, 5.9270e-03 + 4.5925e-04)
  ( 4, 1.9771e-04 + 2.9338e-05)
  ( 6, 8.8541e-06 + 4.8687e-06)
  ( 8, 7.6885e-07 + 5.7900e-07)
  (10, 4.8363e-07 + 6.8699e-07)
  (12, 0.0000e+00 + 0.0000e+00)
  (14, 0.0000e+00 + 0.0000e+00)
};
\addplot[name path=lowerN4, draw=none, forget plot] coordinates {
  (-2, 1.6772e-01 - 1.9573e-03)
  ( 0, 5.8392e-02 - 2.1831e-03)
  ( 2, 5.9270e-03 - 4.5925e-04)
  ( 4, 1.9771e-04 - 2.9338e-05)
  ( 6, 8.8541e-06 - 4.8687e-06)
  ( 8, 7.6885e-07 - 5.7900e-07)
  (10, 4.8363e-07 - 6.8699e-07)
  (12, 0.0000e+00 - 0.0000e+00)
  (14, 0.0000e+00 - 0.0000e+00)
};
\addplot[orange!80, opacity=0.6, forget plot] fill between[of=upperN4 and lowerN4];

\addplot[
  semithick,
  color=purple!80!black,
  mark=x,
  mark options={scale=0.9},
  forget plot
] coordinates {
  (-2, 1.4212e-01)
  ( 0, 3.3649e-02)
  ( 2, 1.4637e-03)
  ( 4, 1.6241e-05)
  ( 6, 1.0629e-08)
  ( 8, 0.0000e+00)
  (10, 0.0000e+00)
};
\addplot[name path=upperN5, draw=none, forget plot] coordinates {
  (-2, 1.4212e-01 + 6.3279e-03)
  ( 0, 3.3649e-02 + 4.8306e-03)
  ( 2, 1.4637e-03 + 3.2889e-04)
  ( 4, 1.6241e-05 + 6.1520e-06)
  ( 6, 1.0629e-08 + 2.6036e-08)
  ( 8, 0.0000e+00 + 0.0000e+00)
  (10, 0.0000e+00 + 0.0000e+00)
};
\addplot[name path=lowerN5, draw=none, forget plot] coordinates {
  (-2, 1.4212e-01 - 6.3279e-03)
  ( 0, 3.3649e-02 - 4.8306e-03)
  ( 2, 1.4637e-03 - 3.2889e-04)
  ( 4, 1.6241e-05 - 6.1520e-06)
  ( 6, 1.0629e-08 - 2.6036e-08)
  ( 8, 0.0000e+00 - 0.0000e+00)
  (10, 0.0000e+00 - 0.0000e+00)
};
\addplot[purple!70, opacity=0.6, forget plot] fill between[of=upperN5 and lowerN5];

\addplot[
  semithick,
  color=blue,
  dashed,
  mark=diamond*,
  mark options={scale=0.9},
  forget plot
]
coordinates {
  (-4, 3.2658e-01)
  (-2, 2.9414e-01)
  ( 0, 2.4782e-01)
  ( 2, 1.8164e-01)
  ( 4, 8.6621e-02)
  ( 6, 1.2633e-02)
  ( 8, 4.7169e-04)
  (10, 6.2999e-06)
  (12, 8.0000e-09)
};

\addlegendimage{line legend, semithick, blue, dashed, mark=diamond*, mark repeat=2,
  mark options={scale=0.9, fill=blue}}
\addlegendentry{{Perfect CSI}$_{(N_R=1)}$}

\addlegendimage{line legend, semithick, blue,
  mark=o, mark repeat=2,
  mark options={scale=0.9, fill=blue}}
\addlegendentry{Transformer$_{(N_R=1)}$}

\addlegendimage{line legend, semithick, red,
  mark=triangle*, mark repeat=2,
  mark options={scale=0.9, fill=red}}
\addlegendentry{Transformer$_{(N_R=2)}$}

\addlegendimage{line legend, semithick, green!70!black,
  mark=square*, mark repeat=2,
  mark options={scale=0.9, fill=green!70!black}}
\addlegendentry{Transformer$_{(N_R=3)}$}

\addlegendimage{line legend, semithick, orange!90!black,
  mark=diamond*, mark repeat=2,
  mark options={scale=0.9, fill=orange!90!black}}
\addlegendentry{Transformer$_{(N_R=4)}$}

\addlegendimage{line legend, semithick, purple!80!black,
  mark=x, mark repeat=2}
\addlegendentry{Transformer$_{(N_R=5)}$}

    \end{axis}
  \end{tikzpicture}
  \caption{Coded BER versus $E_b/N_0$ for $N_R \in \{1,\dots,5\}$ with shaded mean $\pm$ standard deviation regions.}
  \label{fig:results_ber_vs_snr}
\end{figure}
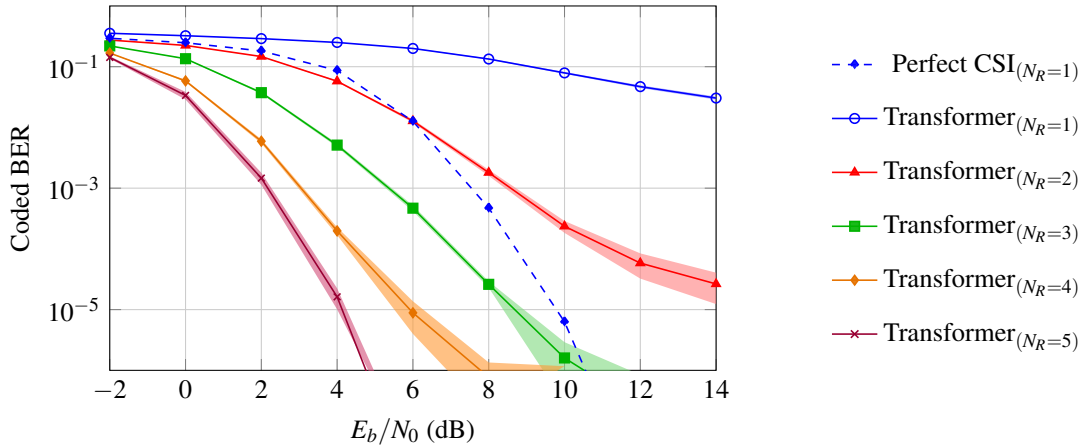

Fig.~\ref{fig:results_ber_vs_snr} reports the coded BER as a function of the normalized $E_b/N_0$ for the proposed pilotless Transformer receiver with $N_R \in \{1,\dots,5\}$ cooperating APs, and for an idealized single-link benchmark with perfect CSI and soft demapping. For each value of $N_R$, the full training and evaluation procedure is repeated ten times with independent random initializations of the neural receiver, and the mean across these runs together with the corresponding standard deviation are shown. Overall, the curves show that fully pilotless decoding is feasible in this realistic ray-traced Wi‑Fi scenario as soon as some spatial diversity is available.

For $N_R = 1$, the Transformer-based receiver does not exhibit a clear waterfall region: the coded BER remains around $10^{-1}$ even at $E_b/N_0 \approx 10$~dB. This is inadequate for a practical Wi‑Fi link and is consistent with the difficulty of learning a reliable channel representation from a single noisy observation without any pilots. In contrast, as soon as $N_R > 1$, the Transformer-based receiver enters a true waterfall region: the BER decreases rapidly with $E_b/N_0$, and adding APs yields an almost horizontal shift of the curves, corresponding to several dB of effective gain at a given target BER. For instance, with $N_R = 2$, pilotless cooperative decoding already provides a useful operating region near BER $10^{-3}$ around $E_b/N_0 \approx 8$~dB, which is comparable to the single-link perfect-CSI benchmark at the same BER. For $N_R = 3$, the same BER is reached at roughly $E_b/N_0 \approx 6$–$7$~dB, corresponding to a gain of a few dB compared to the perfect-CSI single link.

For $N_R = 4$ and $N_R = 5$, the Transformer-based receiver can reliably reach BER values on the order of $10^{-5}$ at moderate SNR. In particular, with $N_R = 4$ the coded BER drops below $10^{-5}$ around $E_b/N_0 \approx 6$~dB, while with $N_R = 5$ similar reliability is achieved already around $E_b/N_0 \approx 4$~dB, and the BER further decreases down to about $10^{-8}$ near $E_b/N_0 \approx 6$~dB. This indicates that, in the presence of a few cooperating APs, fully pilotless operation is sufficient to support highly reliable Wi‑Fi communication for target BERs between $10^{-5}$ and $10^{-6}$. The error bars (mean $\pm$ one standard deviation across independent training runs) are very narrow over most of the operating range, showing that the performance of the learned receiver is robust to random initialization and training noise.

All curves are plotted versus a normalized $E_b/N_0$ that already accounts for the pilot overhead in the perfect-CSI benchmark (i.e., two OFDM pilot columns over 14 symbols). In other words, at a fixed $E_b/N_0$, the pilotless Transformer-based receiver effectively uses more data symbols per frame than the pilot-based receiver, which translates into higher spectral efficiency. Furthermore, for $N_R \geq 2$, the pilotless multi-AP receiver even outperforms the single-link perfect-CSI benchmark over all the considered $E_b/N_0$ range, despite not having access to explicit CSI.

\section{Conclusion}

We have investigated a fully pilotless multi-AP Wi‑Fi receiver based on a self-attention Transformer operating directly on raw complex frequency-domain samples. The proposed architecture outputs bit-wise logits for a standard LDPC decoder without any explicit CSI or pilots.

In a realistic ray-traced indoor scenario with 16‑QAM and rate-$2/3$ LDPC coding, the results show that purely receiver-side pilotless decoding is feasible, provided that some spatial diversity is available. The single-AP case does not reach practical BER levels, which is consistent with the difficulty of inferring a reliable channel from a single noisy observation. In contrast, cooperative configurations with $N_R \geq 2$ exhibit a clear waterfall and operate at coded BERs down to $10^{-5}$–$10^{-6}$ at moderate $E_b/N_0$. Interestingly, they even surpass the perfect-CSI single-link benchmark over a large portion of the considered operating range, despite not using any pilots and benefiting from higher spectral efficiency.

This work being a first step in this direction, future research will aim to confirm these results with multi-user uplinks and explore simpler attention mechanisms, such as axial or structured attention. Such mechanisms have the potential to reduce complexity and help us move towards online, embedded deployments.

\end{document}